\documentstyle[aps,multicol,epsfig,prl]{revtex}

\input epsf
\begin{document}

\title
{Dual Nature of the Electronic Structure of the Stripe Phase}

\author{X. J. Zhou$^{1,2}$, T. Yoshida$^{3}$, S. A. Kellar$^{1}$,
P. V. Bogdanov$^{1}$, E. D. Lu$^{2}$, A. Lanzara$^{1,2}$, M.
Nakamura$^{2}$, T. Noda$^{4}$, T. Kakeshita$^{4}$, H. 
Eisaki$^{4}$, S. Uchida$^{4}$, A. Fujimori$^{3}$, Z. Hussain$^{2}$ 
and Z.-X. Shen$^{1}$} 
\address{ $^{1}$Department of Physics, Applied Physics and Stanford
Synchrotron Radiation Laboratory, Stanford  University,  Stanford,
CA 94305 }
\address{ $^{2}$Advanced Light Source, Lawrence Berkeley National Lab,
Berkeley, CA 94720}
\address{$^{3}$ Department of Physics, University of Tokyo, Bunkyo$-$ku,
Tokyo 113, Japan}
\address{$^{4}$ Department of Superconductivity, University of Tokyo,
 Bunkyo$-$ku, Tokyo 113, Japan}

\date{\today}
\maketitle
\begin{abstract}

High resolution angle$-$resolved photoemission measurements have 
been carried out on (La$_{1.4-x}$Nd$_{0.6}$Sr$_{x}$)CuO$_4$, a 
model system with static stripes, and 
(La$_{1.85}$Sr$_{0.15}$)CuO$_4$, a high temperature superconductor 
(T$_c$=40K) with dynamic stripes. In addition to the straight 
segments near ($\pi$, 0) and (0, $\pi$) antinodal regions, we have 
identified the existence of nodal spectral weight and its 
associated Fermi surface in the electronic structure of both 
systems. The ARPES spectra in the nodal region show well$-$defined 
Fermi cut-off, indicating a metallic character of this 
charge$-$ordered state. This observation of nodal spectral weight, 
together with the straight segments near antinodal regions, 
reveals dual nature of the electronic structure of stripes due to 
the competition of order and disorder. 

\end{abstract}

\begin{multicols}{2}
\narrowtext

The existence and origin of self$-$organized charge stripe and its implication 
to high temperature superconductivity are at the heart of a great debate in 
physics\cite{EmerynZaanen}. Static stripe formation in cuprates was first 
observed in (La$_{2-x-y}$Nd$_y$Sr$_x$)CuO$_4$ (Nd$-$LSCO) system from neutron 
scattering\cite{Tranquada}, with complimentary evidence from other 
techniques\cite{Zimmermann,Imai,Noda,Zhou}. Similar signatures identified in 
(La$_{2-x}$Sr$_x$)CuO$_4$ (LSCO)\cite{Cheong,Mason,Bianconi,Yamada} and other 
high temperature superconductors\cite{Wells,Mook} point to the possible 
existence of stripes in these systems, albeit of dynamic nature. A key issue 
about this new electronic state of matter concerns whether the stripe phase is 
intrinsically metallic or insulating, given its spin and charge 
ordered nature,  and more significantly,  whether it is responsible for high 
temperature superconductivity\cite{Emery,Kivelson}. Understanding the
electronic structure of the stripe phase is a prerequisite for addressing these 
issues and angle$-$resolved photoemission spectroscopy (ARPES) proves to be a 
powerful tool to provide these essential information\cite{Shen}. In this paper, 
we report detailed ARPES results on the electronic structure of 
(La$_{1.4-x}$Nd$_{0.6}$Sr$_x$)CuO$_4$ with static stripes and a related 
superconductor La$_{1.85}$Sr$_{0.15}$CuO$_4$ (T$_c$=40K) with dynamic 
stripes. With high resolution ARPES spectra densely collected under various 
measurement geometries, we have identified the existence of nodal spectral 
weight and its associated Fermi surface in the electronic structure of both 
stripe systems, indicating a metallic nature of this charge$-$ordered state. The 
observation of nodal spectral weight, combined with the straight segments near 
($\pi$,0) and (0,$\pi$) antinodal regions\cite{Zhou}, provides a complete view 
of the dual nature of the electronic structure of the stripe phase, revealing 
the competition of order and disorder in these systems. 

The experiment was carried out at beamline 10.0.1.1 of the Advanced Light 
Source\cite{Zhou}. The angular mode of the Scienta analyzer (SES$-$200) allows 
us to measure an angle of $\sim$14 degrees in parallel (denoted as $\phi$$-$scan
hereafter), corresponding to $\sim$1.1$\pi$ in the momentum
space for the 55eV photon energy we used (the unit of momentum is
defined as 1/{\it a} with {\it a} being the lattice constant). The
momentum resolution is 0.02$\pi$ and the energy resolution is 16$\sim$20 meV. 
To check for possible polarization dependence and  matrix element 
effect\cite{Bansil}, and particularly to map different {\it k}$-$space regions 
of interest, we used various measurement geometries. (1). The sample was 
pre$-$oriented so that the $\phi$$-$scan is along the  crystal axis {\it a} 
(Cu$-$O bonding direction) and the mapping is realized by rotating the sample to 
change the polar angle (denoted as $\theta$$-$scan hereafter) (Fig. 1). (2). The 
sample is oriented so that the $\phi$$-$scan spans the diagonal direction. But 
in this case, the mapping is realized by {\it rotating the analyzer} to change 
the polar angle while keeping the sample fixed (Fig. 2).   In the first 
configuration, the electrical field vector {\bf \overrightarrow{E}} of the 
incident light is nearly normal to the sample surface, while for the second 
configuration, it is parallel to the sample surface. The {\it 
k}$-$space sampling is highly dense; the maps reported here are 
constructed from nearly 10,000 energy distribution curves (EDCs). The (La$_{1.4-
x}$Nd$_{0.6}$Sr$_x$)CuO$_4$ (x=0.10 and 0.15) and (La$_{2-x}$Sr$_x$)CuO$_4$ 
(x=0.15, T$_c$=38K) single crystals were grown using the traveling floating zone 
method\cite{Noda}. The sample was cleaved {\it in situ} in vacuum and measured 
at 15 K with a base pressure of 2$\sim$5$\times$10$^{-11}$ Torr.

Fig. 1 shows the spectral weight of the Nd$-$LSCO (x=0.10 and 0.15)
samples. The data for the x=0.12 sample, taken under the same experimental 
condition,  were previously reported\cite{Zhou}. The integration windows are set 
at 30 meV and 300 meV from the Fermi level, which approximately represent the 
Fermi level feature {\it A(k, E$_F$)} and the momentum distribution function 
{\it n(k)}, respectively, modified by the photoionization matrix 
element\cite{Straub}. The low energy spectral weight for both samples (Fig. 1(a) 
and (c)) is confined near the ($\pi$,0) and (0,$\pi$) regions. Upon increasing 
the energy integration window, the spectral weight extends to the center of the 
Brillouin zone ($\Gamma$ point) (Fig. 1(b) and (d)). In both cases, the spectral 
weight is confined within the boundaries defined by $\mid$k$_x$$\mid$=$\pi$/4 
and $\mid$k$_y$$\mid$=$\pi$/4 (white dashed lines in Fig. 1). It is noted that, 
while the boundary of the spectral weight confinement for the x=0.10 sample is 
very straight, it shows a small wavy character for the x=0.15 sample (Fig. 
1(d)), which is related to another intensity maximum clearly discernable near 
k$_x$=0.5$\pi$. Similar results have also been observed for the x=0.12 
sample\cite{Zhou} and this wavy effect appears to get stronger with doping.

\begin{figure}[t!]
\centerline{\epsfig{figure=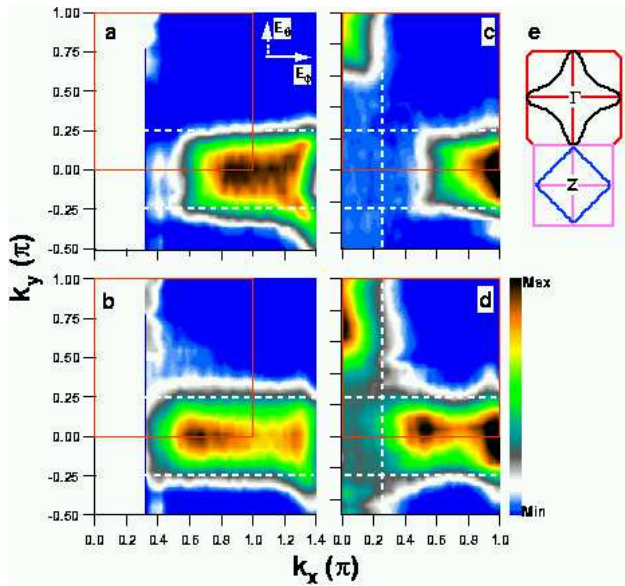,width=8.2cm,clip=}}

\vspace{.2cm} \caption{(color) Spectral weight of Nd$-$LSCO with 
different doping levels, x, integrated within different energy 
windows near the Fermi level. (a). x=0.10, 30meV; (b). x=0.10, 
300meV; (c). x=0.15, 30meV; (d). x=0.15, 300meV. The boundary of 
the spectral weight can be approximated as 
$\mid$k$_x$$\mid$=$\pi$/4 and $\mid$k$_y$$\mid$=$\pi$/4 lines 
(dashed white lines). In the inset of Fig. 1(a) is the in$-$plane 
electrical field component. For comparison, the LDA calculated 
Fermi surface is included in (e) where the top part shows the 
Fermi surface at $k_z$=0 (black line) and the lower part the Fermi 
surface at $k_z=\pi/c$ (blue line)[18].} 
\label{Figure1} 
\end{figure}

The data in Fig. 1 is difficult to reconcile with the band structure 
calculations where the LDA calculated Fermi surface runs more or less diagonally 
([1,-1] direction) (see Fig. 1(e))\cite{Xu}. We have done a matrix element 
simulation for this measurement geometry and found that the spectral weight near 
($\pi$/2,$\pi$/2) nodal region is suppressed compared with the ($\pi$,0) and 
(0,$\pi$) antinodal regions. Nevertheless, it is still impossible to 
reproduce the observed spectral weight confinement with straight boundaries 
parallel to [1,0] or [0,1] direction, even by considering matrix element 
effects as well as the $k_z$ effect.  On the other hand, a simple stripe picture 
seems to capture the main characteristics of the data in Fig. 1 if one considers 
the low energy signal as mainly being from metallic 
stripes\cite{Zhou,Salkola,Tohyama,Fleck,Markiewicz}. The momentum distribution 
function {\it n(k)} (Fig. 1(b) and (d)) then suggests two sets of Fermi 
surfaces, defined by the  $\mid$k$_x$$\mid$=$\pi$/4 and 
$\mid$k$_y$$\mid$=$\pi$/4 lines, which can be understood as a superposition of 
two one$-$dimensional (1D) Fermi surfaces originating from two perpendicularly 
orientated stripe domains. The $\pi$/4 boundary lines are related to the fact 
that the stripes are approximately quarter$-$filled in the charge 
stripes\cite{Tranquada}. 

However, the above rigid stripe model also leaves many questions unanswered. A 
prominent one relates to the fast dispersion seen along (0,0) to ($\pi$,0) 
direction within the 200 meV of the Fermi level, which implies a charge motion  
perpendicular to vertical stripes\cite{Zhou}.  The concomitant carrier motion 
along stripes and transverse to stripes points to the possible existence of a 
nodal state along the [1,1] diagonal direction which has zero superconducting 
gap in {\it d}$-$wave pairing symmetry. However, as seen from Fig. 1, as well as 
previous measurements for underdoped samples\cite{Ino,Zhou}, there is little 
spectral weight observed near the nodal region. 

\begin{figure}[t!]
\centerline{\epsfig{figure=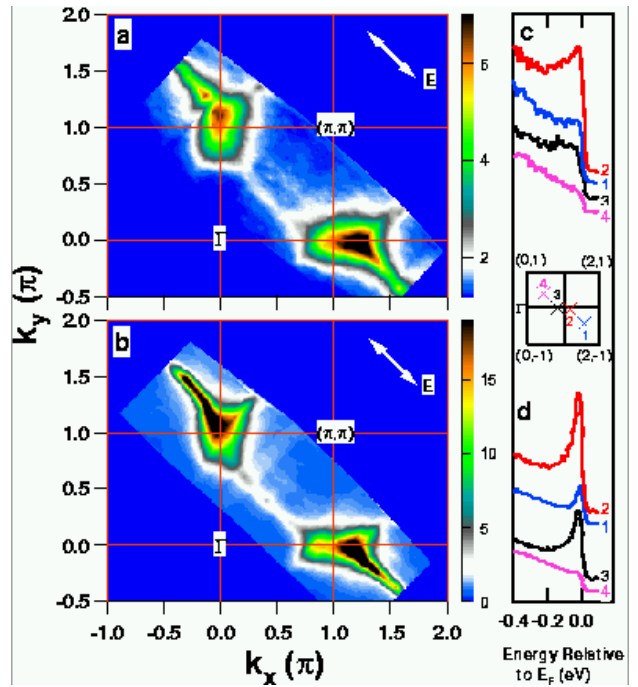,width=8.2cm,clip=}}

\vspace{.2cm} \caption{(color) Spectral weight of Nd$-$LSCO (x=0.15) (a) and 
LSCO (x=0.15) (b) integrated within 30 meV energy window near the Fermi
level. Typical EDC spectra near the nodal and the ($\pi$, 0) regions, 
as indicated in the in$-$between inset, are shown in (c) and (d) for Nd$-$LSCO 
and LSCO, respectively. The in$-$plane electrical field is also indicated which 
was fixed during the measurement.} \label{Figure2}
\end{figure}

To gain more insight into the issue of nodal state, we focused our measurements 
on the nodal regions by covering along the diagonal direction (Fig. 2).  We also 
extended the measurement to the second zone by considering possible different 
matrix element for different zones,  as well as sharpening of spectral line in 
the second zone due to the negative dispersion velocity involved\cite{Hansen}. 
In order to enhance the signal from the nodal region, the in$-$plane electric 
field is maximized by fixing the {\bf \overrightarrow{E}} vector parallel to the 
sample surface; the measurement was fulfilled by rotating the analyzer, thus 
avoiding the complications of varying polarization. Our matrix element 
simulations show spectral weight enhancement in the nodal region for this 
measurement geometry.  Fig. 2 shows the low energy excitations for both 
Nd$-$LSCO (x=0.15) and La$_{1.85}$Sr$_{0.15}$CuO$_{4}$ samples, measured under 
such a geometry. One sees the similar straight segments near (0,$\pi$) and 
($\pi$,0) as seen in Fig. 1 even though the {\it a-b} plane of the samples were 
rotated 45 degrees with respect to each other.  More importantly, it is possible 
to see the spectral weight near {\it d}$-$wave nodal region for both samples, 
although the signal in the first zone is weaker than that in the second 
zone. The EDC spectra near the nodal regions show a well$-$defined Fermi 
cut$-$off (Fig. 2(c) and (d)), indicating the metallic nature of these systems.

As seen from Fig. 1 and 2, there are two aspects involved in the electronic 
structure of the stripe phase. The first is the straight segment near the 
($\pi$,0) and (0,$\pi$) antinodal regions. This feature is very robust as it 
is seen under different measurement geometries, which appears to be a measure of 
the ordered nature of stripes. The second feature is the nodal state seen in 
Fig. 2, which is sensitive to doping and measurement geometry. This feature is 
likely a measure of the deviation from ideal stripe case, which may be due to 
disorder or dynamic fluctuations. We note here that we have observed similar two 
features in Nd$-$LSCO with x=0.10 and in LSCO with doping level as low as 
x=0.07\cite{ZhouToBe}. The nodal signal appears to get stronger with increasing 
Sr doping for both Nd$-$LSCO and LSCO, and for a given Sr doping level, it is 
stronger in Nd$-$free LSCO than in Nd$-$LSCO samples\cite{ZhouToBe}. As seen 
from Fig. 2(a) and (b),  near the ($\pi$,0) region in the second zone, the 
maximum intensity contour extends all the way from the nodal region and 
intersects with the ($\pi$,0)$-$(2$\pi$,0) line, suggesting an electron$-$like 
Fermi surface. Interestingly, this Fermi surface is reminiscent of that from the 
LDA calculation (Fig. 1(e))\cite{Xu}. 
 
A complete description of the electronic structure of the stripe phase needs to 
take both of the above two aspects into account: the straight segment near the 
antinodal region and the spectral weight near the nodal region with its 
associated Fermi surface.   Fig. 3 highlights these two features in the first 
Brillouin zone for the Nd$-$LSCO (x=0.15) and LSCO (x=0.15) samples, which are 
also schematically depicted in Fig. 3(c) (upper$-$left panel). While it seems to 
be straightforward to associate the straight segments with stripes because of 
their 1D nature\cite{Salkola}, the detection of spectral weight near the nodal 
region in the {\it static} stripe phase poses a new challenge to our 
understanding of this charge ordered state because the nodal spectral weight is 
usually expected to be suppressed in a simple stripe 
picture\cite{Tohyama,Fleck}.  The experimental question regarding the origin is 
whether they originate from another distinct phase or they are intrinsic 
properties of the same stripe phase. In the case of phase separation, this would 
mean that, besides stripes, there is another non$-$stripe metallic phase with a 
much higher carrier concentration, as estimated from the Luttinger volume of the 
diamond$-$shaped Fermi surface (Fig. 3). As far as we know, there is little 
evidence of such a phase separation in Nd$-$LSCO and LSCO systems at the doping 
level discussed here\cite{Ino},  although it has been observed in a related 
La$_2$CuO$_{4+\delta}$ system\cite{Wells}. 

\begin{figure}[t!]
\centerline{\epsfig{figure=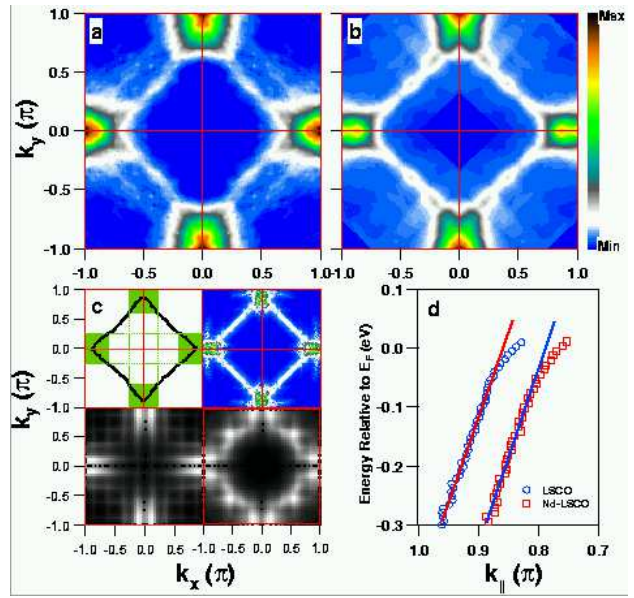,width=8.2cm,clip=}}

\vspace{.2cm} \caption{(color) Measured low energy (30meV 
integration) spectral weight of Nd$-$LSCO (x=0.15) (a) and LSCO 
(x=0.15)(b), as obtained by symmetrizing the first zone data of 
Fig. 2. The observed two features are schematically illustrated in 
(c) (upper$-$left panel): diamond$-$shaped nodal Fermi surface 
(black line) and 1D spectral confinements near ($\pi$,0) and 
(0,$\pi$) regions. The spectral weight patterns calculated from 
stripe fluctuation (upper-right panel)[19], from the site-centered 
stripe (lower$-$left panel) and bond$-$centered stripe 
(lower$-$right panel) [26] are also included in (c) for 
comparison. Fig. 3(d) shows the dispersion along the nodal 
direction (in the second zone) for the Nd$-$LSCO and LSCO samples; 
a slope breakdown in the dispersion can be seen for both cases at 
nearly -50meV. The dispersion for Nd$-$LSCO is horizontally offset 
for clarity, with the two solid lines as guide to the eye. } 
\label{Figure3} 
\end{figure}

The detection of nodal spectral intensity in the stripe system provides a clear 
distinction of the stripe physics from the ordinary 1D charge motion in a rigid 
one$-$dimensional system. In the stripe context, the nodal Fermi surface may 
arise from disorder or fluctuation of stripes where the holes leak into the 
antiferromagnetic region\cite{Salkola,Markiewicz}. Here disorder also induces 
the effect that the antiferromagnetic region may not be fully gapped when it 
becomes very narrow.  The measured spectral weight in Fig. 3 for Nd$-$LSCO 
and LSCO is similar to the one calculated based on such a disordered stripe 
picture (Fig. 3(c), upper$-$right panel)\cite{Salkola}. This also seems to be 
consistent with the trend that in LSCO the nodal spectral 
weight is more intense than that in Nd-LSCO because the stripes in the former 
are dynamic while they are static in the latter. Note that, in this picture, the 
nodal Fermi surface is actually a superposition of Fermi surface features from 
two perpendicular stripe domains; for an individual stripe domain, this Fermi 
surface feature can be discrete\cite{Salkola,Markiewicz}. 

An alternative scenario to understand the two features in the stripe context is 
a possible coexistence of site$-$ and bond$-$centered stripes\cite{Zacher}. Both 
types of stripes are compatible with neutron experiments\cite{Tranquada}, and 
are close in energy as indicated by various calculations\cite{Scalapino}. As 
shown in Fig. 3(c), the calculated {\it A(k,E$_F$)} patterns for the 
site$-$centered (lower$-$left panel) and bond$-$centered stripes (lower$-$right 
panel)\cite{Zacher} bear a clear resemblance to the data in Fig. 1 and Fig. 2, 
respectively. It is therefore tempting to associate the 1D straight segment to 
site$-$centered stripes and the nodal Fermi surface feature to bond$-$centered 
stripes. Since these two types of stripes are different in their wave functions, 
it may also help explain why they show different behaviors under different 
measurement conditions. Moreover, the sudden slope change (or "kink") of the 
dispersion along the nodal direction for both Nd$-$LSCO and LSCO (Fig. 3(d)) is 
also consistent with this scenario\cite{Fleck,Zacher}.  If this picture proves 
to be true, it would imply that, with increasing doping, bond$-$centered stripes 
are produced at the expense of site$-$centered stripes, and more bond$-$centered 
stripes may be generated as the stripes become more dynamic (Fig. 3).  This 
seems to further suggest that the bond$-$centered stripes are more favorable for 
superconductivity than the site$-$centered stripes, a possibility remains to be 
investigated further. We note that the above scenarios can be closely related to 
each other. Stripe disorder or fluctuation may naturally give rise to hole$-
$rich and hole$-$poor regions as in phase separation case, particularly with the 
possible existence of stripe dislocations in the system. In the case of stripe 
fluctuation, the randomness in the stripe separations may necessarily give rise 
to a mixture of site$-$centered and bond$-$centered stripes.

We thank J. Denlinger and J. Bozek for technical support,  S. Kivelson, J. 
Tranquda, A. Balatsky, W. Hanke, M. Zacher and M. Salkola for helpful 
discussions. The experiment was performed at the Advanced Light Source of the 
Lawrence Berkeley National Laboratory, which is operated by the DOE's Office of 
Basic Energy Science, Division of Material Science, with contract DE-AC03-
76SF00098. The division also provided support for the work at SSRL. The work at 
Stanford was supported by NSF grant through the Stanford MRSEC grant and NSF 
grant DMR-9705210.

\end{multicols}
\end{document}